\documentclass[aps,prd,superscriptaddress,twocolumn,nofootinbib,preprintnumbers]{revtex4}
\pdfoutput=1

\usepackage{color}
\usepackage{graphicx}
\usepackage[space]{grffile}

\usepackage{verbatim}
\usepackage{amsmath}
\usepackage{amssymb}
\usepackage[caption=false]{subfig}
\usepackage{url}
\usepackage{bbold}
\usepackage{slashed}
\usepackage{epstopdf}

\usepackage{multirow}
\usepackage{threeparttable}
\usepackage{paralist}

\usepackage{color}
\definecolor{darkgreen}{rgb}{0,0.5,0}
\definecolor{violet}{rgb}{0.5,0.5,1}

\newcommand{\MeV}{\text{MeV}}
\newcommand{\GeV}{\text{GeV}}

\DeclareRobustCommand{\Sec}[1]{Sec.~\ref{#1}}

\DeclareRobustCommand{\App}[1]{App.~\ref{#1}}
\DeclareRobustCommand{\Tab}[1]{Table~\ref{#1}}

\DeclareRobustCommand{\Fig}[1]{Fig.~\ref{#1}}

\DeclareRobustCommand{\Eq}[1]{Eq.~(\ref{#1})}

\DeclareRobustCommand{\Ref}[1]{Ref.~\cite{#1}}

\newcommand{\be}{\begin{equation}}
\newcommand{\ee}{\end{equation}}
\newcommand{\bea}{\begin{eqnarray}}
\newcommand{\eea}{\end{eqnarray}}
\newcommand{\bi}{\begin{itemize}}
\newcommand{\ei}{\end{itemize}}

\usepackage{xspace}

\usepackage[colorlinks=true
,urlcolor=blue
,anchorcolor=blue
,citecolor=blue
,filecolor=blue
,linkcolor=blue
,menucolor=blue
,linktocpage=true
,pdfproducer=medialab
,pdfa=true
]{hyperref}

\begin{document}

\title{Boosted Dark Matter at Neutrino Experiments}

\author{Lina Necib}
\email{lnecib@mit.edu}

\affiliation{Center for Theoretical Physics, Massachusetts Institute of Technology, Cambridge, MA 02139, USA}

\author{Jarrett Moon}

\affiliation{Laboratory for Nuclear Science, Massachusetts Institute of Technology, Cambridge, MA 02139, USA}

\author{Taritree Wongjirad}
\affiliation{Laboratory for Nuclear Science, Massachusetts Institute of Technology, Cambridge, MA 02139, USA}

\author{Janet M. Conrad}
\affiliation{Laboratory for Nuclear Science, Massachusetts Institute of Technology, Cambridge, MA 02139, USA}

\preprint{MIT-CTP/4840} 
\begin{abstract}
Current and future neutrino experiments can be used to discover dark matter, not only in searches for dark matter annihilating to neutrinos, but also in scenarios where dark matter itself scatters off Standard Model particles in the detector. 
In this work, we study the sensitivity of different neutrino detectors to a class of models called boosted dark matter, in which a subdominant component of a dark sector acquires a large Lorentz boost today through annihilation of a dominant component in a dark matter-dense region, such as the galactic Center or dwarf spheroidal galaxies. 
This analysis focuses on the sensitivity of different neutrino detectors, specifically the Cherenkov-based Super-K and the future argon-based DUNE to boosted dark matter that scatters off electrons.
 We  study the dependence of the expected limits on the experimental features, such as energy threshold, volume and exposure in the limit of constant scattering amplitude. 
We highlight experiment-specific features that enable current and future neutrino experiments to be a powerful tool in finding signatures of boosted dark matter. 
\end{abstract}

\maketitle

\section{Introduction}

Gravitational evidence for dark matter (DM) is overwhelming \cite{Zwicky:1933gu,Rubin:1980zd,Clowe:2006xq}, but all nongravitational means of DM detection have not yet resulted in a definitive discovery. It is therefore essential to expand DM searches to encompass as many possible DM signals. Previous work \cite{Agashe:2014yua} has proposed a new class of DM models called boosted dark matter (BDM) with novel experimental signatures at neutrino experiments. BDM search strategies are complementary to existing indirect detection searches for DM at neutrino detectors. 

BDM expands the weakly interacting massive particle (WIMP) paradigm to a multicomponent dark sector that includes a component with a large Lorentz boost obtained today due to decay or annihilation of another dark particle at a location dense with DM.
In this class of models, the boosted component can scatter off standard model (SM) particles similarly to neutrinos, and can thus be detected at neutrino experiments. Various extensions built on the BDM model \cite{Berger:2014sqa,Kong:2014mia,Cherry:2015oca,Kopp:2015bfa} have studied the potential reach at large volume neutrino detectors and even direct detection experiments.

\begin{figure}[t]
\begin{center}
\includegraphics[scale=0.6, trim = 0 0 0 0]{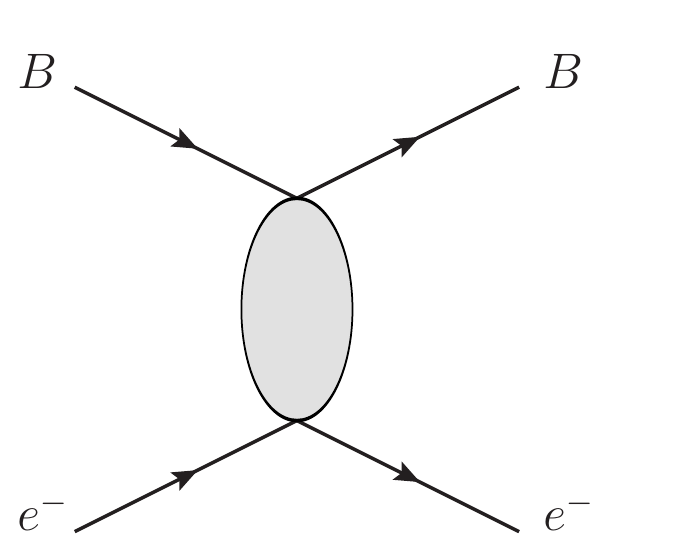}\hspace{2pc}%
\caption{\label{fig:BeBe} Scattering process of BDM $B$ off of electrons.  }
\end{center}
\end{figure}

In this paper, we present BDM searches assuming a constant scattering amplitude, which highlight the reach of different neutrino technologies with different experimental features, and in particular electron energy thresholds. Focusing on scenarios in which BDM scatters off electrons (and leaving scattering off protons to future work \cite{josh}) the scattering process of interest, shown in \Fig{fig:BeBe}, is 
\be
B ~e^- \rightarrow B~ e^-,
\ee
where $B$ is a subdominant DM component with a Lorentz boost due to the annihilation of another heavier dominant state $A$ 
\be
A \overline{A} \rightarrow B \overline{B}.
\ee
as shown in \Fig{fig:feynman}.

We present the potential reach for two searches for BDM, one where the boosted particle $B$ originates at the galactic Center (GC) and one where $B$ originates at dwarf galaxies (dSphs). Although dSphs are a great source for DM since they are low in astrophysical backgrounds, their DM density is lower than that of the GC, so we perform a stacked analysis to increase statistics and improve sensitivity.

We take advantage of $B$'s large Lorentz boost in reducing background as the emitted electrons scatter in the forward direction and therefore point to the origin of the BDM particle. This is different from the omnidirectional atmospheric neutrino background, dominated by the charged current processes 
\begin{eqnarray}
 \nu_e ~n \rightarrow e^- ~p, \\
 \overline{\nu_e}~ p \rightarrow e^+ ~n.
\end{eqnarray}

Experiments of particular interest are Cherenkov detectors like  Super-Kamiokande (Super-K) \cite{Fukuda:2002uc} and Hyper-Kamiokande (Hyper-K) \cite{Lodovico:2015yii}, and liquid argon time projection chambers (LArTPCs) like the upcoming Deep Underground Neutrino Experiment (DUNE) \cite{Acciarri:2015uup}.
Argon-based detectors utilize a new technology that has not previously been thoroughly investigated within the context of DM searches. We explore LArTPCs' excellent angular resolution and particle identification in this paper, and emphasize the discrimination power of LArTPC experiments even with smaller volumes than their Cherenkov counterparts. We show the overall sensitivity of Super-K, Hyper-K and DUNE in setting limits on the DM-SM scattering cross section for the case of annihilation from another heavier component $A$. The decay case can be worked out in a similar fashion.

The rest of this paper is organized as follows: In Sec. \ref{sec:model}, we introduce a simplified parametrization that captures BDM's main features, and set up the framework to relate the expected number of detected events to the general properties of BDM. We then study event selection in Sec. \ref{sec:detection} and background rejection in Sec. \ref{sec:background} for the Cherenkov and argon-based technologies. We finally show the experimental reach at current and future neutrino experiments to BDM originating in the GC in Sec. \ref{sec:dune} and in dSphs in \Sec{sec:dSphs_analysis}, and conclude in Sec. \ref{sec:conclusion}.

\section{Boosted Dark Matter}
\label{sec:model}

\subsection{Features of Boosted Dark Matter}

One of the most studied paradigms of DM is that of WIMPs in which DM is a single cold thermal particle that froze out early in the Universe's history. Various detection methods have been used to search for WIMP DM: direct detection in which nonrelativistic DM particles scatter off heavy nuclei \cite{Jungman:1995df,Akerib:2013tjd,PhysRevLett.112.041302,Akerib:2015rjg}, and indirect detection in which SM particles resulting from DM annihilation/decay are detected (see for example,  \cite{Bergstrom:2013jra,Essig:2013goa,Daylan:2014rsa,Ackermann:2015zua}). Indirect detection signals originate in DM-dense regions, two of which are  the GC and dSphs. 

\begin{figure}[t]
\begin{center}
\includegraphics[scale=0.6, trim = 0 0 0 0]{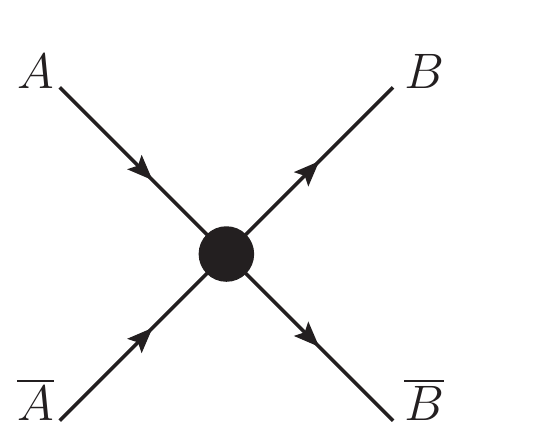}\hspace{2pc}%
\caption{\label{fig:feynman}  Annihilation process that produces $B$ with a Lorentz boost.}
\end{center}
\end{figure}

BDM is a class of multicomponent models in which a component of the dark sector has acquired a Lorentz boost today. 
 Let the DM sector be composed of a dominant component $A$ and a subdominant component $B$.\footnote{$A$ and $B$ can be the same particle as in the case of a $Z_3$ symmetry for example \cite{Agashe:2004ci,Agashe:2004bm,Ma:2007gq,Walker:2009ei,Walker:2009en,DEramo:2010ep}, and $A$ can correspond to more than one particle in the case of a more complex dark sector.}
\begin{itemize}
\item The particle $B$ is boosted due to either annihilation or decay of a second state $A$, as shown in \Fig{fig:feynman}. Other processes that would boost the $B$ particle can be easily derived from the subsequent formalism, such as semiannihilation $A A \rightarrow B \phi$ \cite{DEramo:2010ep} for example, with the energy of $B$ satisfying $E_B \gg m_B$. 
\item The boosted particle $B$ interacts with the SM through a scattering process. In this work, we focus on the case of $B$ scattering off electrons $B e^- \rightarrow B e^-$, as in \Ref{Agashe:2014yua}. We leave the case of $B$ scattering off protons \cite{Berger:2014sqa,Kong:2014mia} to future work \cite{josh}.\footnote{Proton scattering is more important for scenarios where DM, and in this case $A$,  is captured in the Sun. This case depends on the capture scenario rather than the initial DM density, and therefore it is not incorporated in this work.}
\end{itemize}

Searching for BDM therefore involves a hybrid approach, as one would \emph{directly} detect the $B$ particle scattering off SM particles, and at the same time \emph{indirectly} detect the $A$ component. 
In the following , we present a simplified parametrization of BDM in order to compare the reach of different neutrino detector technologies.

\subsection{Flux of Boosted Dark Matter from Annihilation} \label{sec:flux}

The flux of $B$ produced in $A$ annihilation (see Fig. \ref{fig:feynman}) within a region of interest (ROI) of a particular source is 
\be \label{eq:fluxann}
\frac{d\Phi^{\text{ROI}}_{\text{ann}} }{d \Omega dE_B}  = \frac{ j_{\text{ann}}(\Omega) }{8 \pi m_A^2}  \langle \sigma_{ \overline{A} A  \rightarrow \overline{B} B} v\rangle \frac{dN_B}{d E_B}.
\ee
The annihilation J-factor $j_\text{ann}$ is obtained by integrating over the DM density squared along the line of sight at a particular position in the sky,
\be
j_{\text{ann}}(\Omega) = \int_{\text{l.o.s}} ds~ \rho(s) ^2.
\ee
The thermally averaged cross section $\langle \sigma_{\overline{A} A \rightarrow \overline{B} B} v \rangle$ is the annihilation cross section of the process that produces the $B$ particles, taken as a reference to be equal to the thermal cross section $\langle \sigma_{\overline{A} A \rightarrow \overline{B} B} v \rangle = 3 \times 10^{-26} \text{cm}^3/\text{sec}$. Any deviation is an overall rescaling of the flux. 

As was previously argued in \Ref{Agashe:2014yua}, the optimal choice of ROI for the GC analysis is $\approx 10^\circ$ around the GC for the case of annihilation.\footnotemark  ~We therefore adopt the same ROI in this analysis.
We define $J_\text{ann}$ as the integrated J-factor $j_\text{ann} (\Omega)$ over a patch of the sky, assuming an NFW profile \cite{Navarro:1995iw}, as 
\be \label{eq:jann}
J_{\text{ann}}^{10^\circ} = \int d\Omega ~j_{\text{ann}}(\Omega) \stackrel{10^\circ }{=} 1.3 \times 10^{21} \text{GeV}^2/\text{cm}^5 ,
\ee
where the numerical value corresponds to a cone of half angle $10^\circ$ around the GC \cite{Cirelli:2010xx}.\\
The spectrum of $B$ is $dN_B/dE_B$, which in the case of the $A \overline{A} \rightarrow B \overline{B}$ process is
\be
\frac{dN_B}{dE_B} = 2~ \delta (E_B - m_A).
\ee
Therefore, the integrated flux over a patch of the sky is
\be 
\Phi^{\text{GC}}_{\text{ann}} =  \frac{J_{\text{ann}}^{10^\circ} }{4 \pi m_{A}^2} \langle \sigma_{ \overline{A} A  \rightarrow \overline{B} B} v\rangle.
\ee
The numerical values of the flux of DM integrated over the whole sky and over a cone of half angle $10^\circ$ for $\overline{A} A \rightarrow \overline{B} B$ are 
\begin{eqnarray}
\Phi^{\text{GC}}_{\text{ann}}   &=&49.6 \times 10^{-8} ~\text{cm}^{-2} ~\text{sec}^{-1} \left( \frac{20 ~\text{GeV}}{m_A}\right)^2 \nonumber \\
&&\times \left( \frac{\langle \sigma_{\overline{A} A \rightarrow \overline{B} B} v \rangle }{ 3 \times 10^{-26} ~\text{cm}^3/\text{sec}}\right). \\
 \label{eq:phiGCann10}
\Phi^{\text{GC},10^\circ}_{\text{ann}}   &=& 4.7 \times 10^{-8} ~\text{cm}^{-2} ~\text{sec}^{-1} \left( \frac{20 ~\text{GeV}}{m_A}\right)^2 \nonumber \\
&& \times \left( \frac{\langle \sigma_{\overline{A} A \rightarrow \overline{B} B} v \rangle }{ 3 \times 10^{-26} ~\text{cm}^3/\text{sec}}\right).
\end{eqnarray}

 \footnotetext{The value of the optimal opening angle for decay ($A \rightarrow B \overline{B}$) cannot be taken as $10^\circ$ without a proper analysis. The initial value of the opening angle depends largely on the DM distribution. The fact that annihilation signals scale as the DM density squared while decay signals scale linearly with DM density means that DM will be less localized near the center, and that leads to a larger optimal choice of ROI.}

\subsection{Implications of Forward Scattering} \label{sec:detection}

\begin{figure}[t]
\begin{center}
\includegraphics[width=15pc, trim = 0 0 0 0]{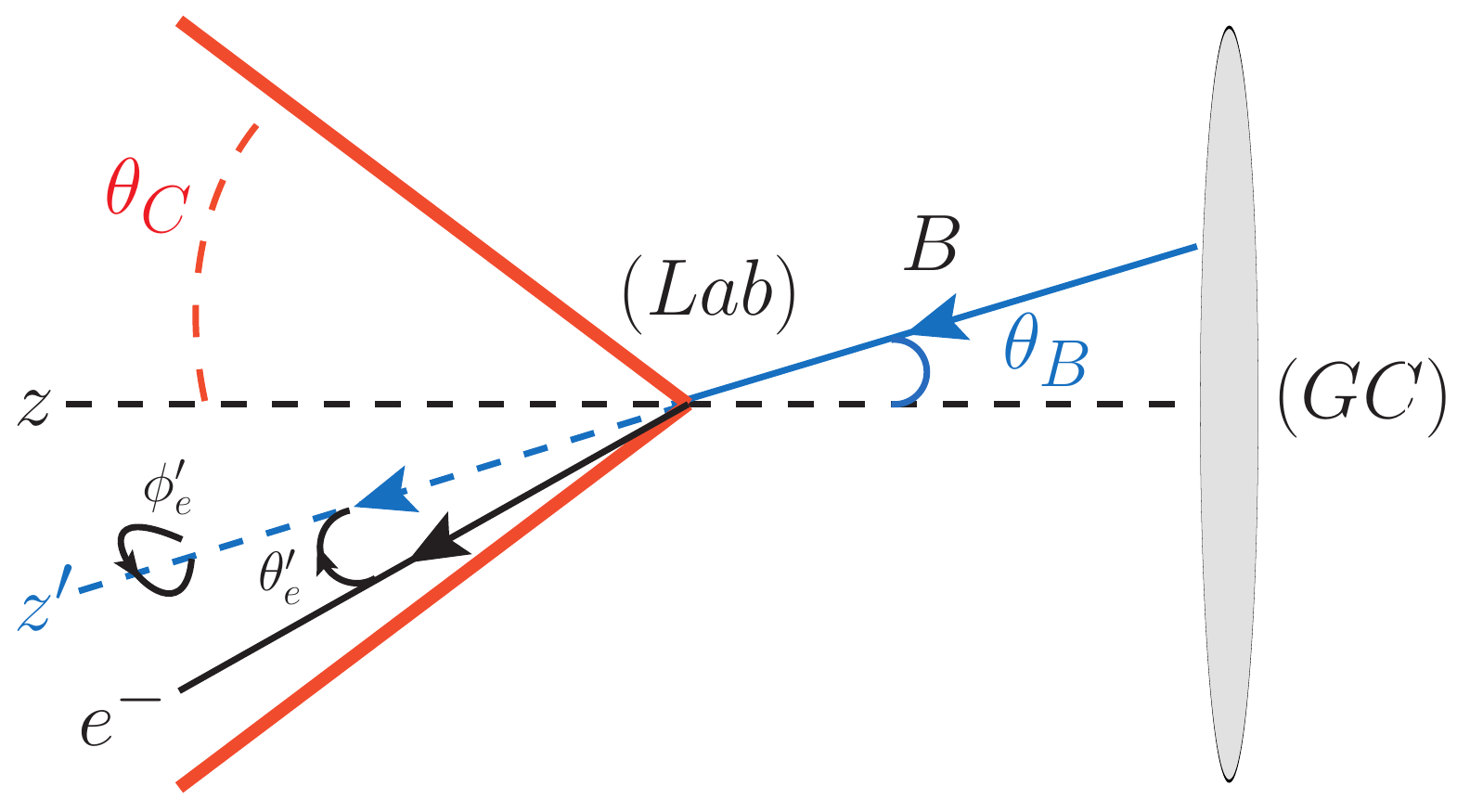}\hspace{2pc}%
\caption{\label{fig:cone} Geometry of a search cone for incoming $B$ particles originating at the GC and scattering off electrons at a neutrino experiment \cite{Agashe:2014yua}. }
\end{center}
\end{figure}

In the energy range of $\mathcal{O} (10 \text{ MeV})  - \mathcal{O} (100~\text{GeV})$, the dominant background for any neutrinolike signal is atmospheric neutrinos \cite{Gaisser:2002jj,Battistoni:2002ew,Dziomba}.\footnote{Solar neutrinos dominate below energies of $30$ MeV. Although we know the location of the Sun and can thereby veto solar neutrinos, we avoid this parameter space in order to be conservative as it is hard to estimate the ability of photomultipliers to trigger on events with such low energies.} The key aspect in discriminating the background, which is omnidirectional, from the signal, which originates at a location dense in DM, is adopting a search cone strategy. As shown in \Fig{fig:cone}, we veto all electrons that are emitted at an angle larger than $\theta_C$ around a particular source. This strategy takes advantage of forward scattering of the electron, emitted in the same direction as the incoming $B$. 

As was computed in Ref. \cite{Agashe:2014yua}, the expected number of electron events $N_\text{signal}^{\theta_C}$ is obtained by convolving the initial DM distribution over the electron scattering angle of the $B e^- \rightarrow B e^-$ process, such that the emitted electron is scattered at angles smaller than $\theta_C$ around a particular source. 
\begin{eqnarray} \label{eq:convolution}
N^{\theta_C}_{\rm{signal}}&=& \Delta T N_{\rm{target}} \nonumber \\
& \times& \int_{\theta_B} d \theta_B \left(f_{B} (\theta_B) \otimes \frac{d \sigma_{B e^- \rightarrow B e^-}}{d \theta_e'} \right)\Big|_{\theta_e' < \theta_C} 
\end{eqnarray} 
where $\Delta T$ is the exposure time, and $N_\text{target}$ is the number of target electrons in the experiment considered. The angle $\theta_B$ is the polar angle of $B$ with respect to the source (GC or dSphs). The angle $\theta'_e$ is the polar angle of $e^-$ with respect to the incoming direction of the $B$ (see \Fig{fig:cone}). $f_B (\theta_B)$ is the flux of the incoming $B$ particles as a function of the polar angle, integrated over the azimuthal angle. 
For a particular source, the total flux is related to $f_B$ by 
\be
\Phi_B^{\alpha} = \int_{0}^{\alpha} f_B (\theta_B) d\theta_B.
\ee
 This is equal to \Eq{eq:phiGCann10} when $\alpha = 10^\circ$.

As we show in \App{app:forward}, in the limit where the energy of the BDM particle is much higher than the electron mass ($E_B \gg m_e$), highly boosted DM (with a Lorentz boost factor $\gamma_B \gg 1$) scatters off electrons which are then emitted in the forward direction ($\theta_e' = 0$). We can therefore use the electron scattering angle to infer the BDM's origin. In this limit, the convolution of \Eq{eq:convolution} can be simplified as 
\be \label{eq:Nevents_factorized}
N^{\theta_C}_{\rm{signal}} =  \Delta T \times N_{\text{target}} \times \Phi_B^{\theta_C} \times \sigma_{B e^- \rightarrow B e^-}^\text{measured}.
\ee

It is important to note that the cross section $ \sigma_{B e^- \rightarrow B e^-}^\text{measured}$, hereafter labeled $\mathcal{I}$, is not the total cross section, but rather the \textit{measured} one, as the energy threshold of the experiment introduces an energy cutoff.

We write the measured cross section $\mathcal{I}$ as a function of the energy threshold $E_\text{thresh}$ in order to facilitate the comparison among experiments with different characteristics. 
Assuming that the limiting experimental factor is the energy threshold rather than the angular resolution, and this is a good approximation that follows from scatterings being in the forward direction and the excellent angular resolution of neutrino experiments, we write the measured cross section as a function of the measured energy of the emitted electron $E_e$
\be \label{eq:integral}
\mathcal{I} (E_\text{thresh})= \int_{E_\text{thresh}}^{E_\text{max}} dE_e \frac{d \sigma_{B e^- \rightarrow B e^-}}{d E_e}.
\ee
The upper limit of integration is
\begin{eqnarray}
E_\text{max} &=& m_e \frac{(E_B + m_e)^2 + E_B^2 - m_B^2}{(E_B + m_e)^2 - E_B^2 + m_B^2}, \label{eq:emax}
\end{eqnarray}
which is the maximum allowed by the kinematics of the scattering process.

 \subsection{Constant Amplitude Limit} \label{sec:constant}

 In order to compare the reach of different experiments, we extract the dependence on the energy threshold while assuming a constant scattering amplitude. This simplifies the parameter space in order to better illustrate the reach of different experiments.

Let $\sigma_0$ be the total cross section for the process $B e^- \rightarrow B e^-$,
\be \label{eq:sigma0}
\sigma_0 = \int_0^{E_\text{max}} dE_e \frac{d \sigma_{B e^- \rightarrow B e^-}}{d E_e}.
\ee
If we assume a flat amplitude $|\mathcal{M}|^2 = $ constant, we can then relate $\mathcal{I}$ defined in \Eq{eq:integral} with $\sigma_0$ defined in \Eq{eq:sigma0} by
\be
\mathcal{I} (E_\text{thresh})= \sigma_0 \left( 1 - \frac{ E_\text{thresh}}{E_\text{max} }\right) . 
\ee
Below, we estimate limits on the quantity $\sigma_0$.
The expected number of events given by \Eq{eq:Nevents_factorized} is 
\be
N^{\theta_C}_{\rm{signal}} = \Delta T \, N_{\rm{target}} ~\Phi_B^{\theta_C}~  \sigma_0 \left( 1 -  \frac{  E_\text{thresh}}{  E_\text{max} } \right) . \label{eq:signal}
\ee

\section{Event Selection}
\label{sec:detection}

The backgrounds to the signal process $B e^- \rightarrow B e^-$ are all processes in which an electron in the appropriate energy range is emitted from neutrino-induced scatterings. The processes with the highest cross sections are charged current neutrino scatterings $\nu_e + n \rightarrow e^- + p$ and $\overline{\nu_e} + p \rightarrow e^+ +  n$. For the energies of interest in $\mathcal{O} (10 ~\MeV) - \mathcal{O} (100 ~\GeV)$, the dominant background is atmospheric neutrinos. 
Neutrinos scattering in detectors produce both electrons and muons while the signal is present only in electron events. Therefore, an important feature of this BDM model is an excess in the electron channel over the muon channel. 
We now study the features of the signal that are used to discriminate against the background in Cherenkov and LArTPCs detectors separately.

\subsection{Cherenkov Detectors: Super-K}

We study Super-K as an example of Cherenkov detectors in this analysis. Super-Kamiokande is a large underground water Cherenkov detector, with a fiducial volume of 22.5 kton of ultrapure water. It has collected over 10 years of atmospheric data, which would be the target data set for this analysis \cite{Fukuda:1998mi,Ashie:2005ik,Wendell:2010md}.

The atmospheric neutrino backgrounds, as well as signal events in Super-K, are single-ring electrons, detected with the following properties.
\begin{itemize}
 \item \textit{Energy range}: for the electron to be detected in a Cherenkov experiment, the electron energy $E_e$ has to be above the Cherenkov limit $\gamma_\text{water} m_e$, with $\gamma_\text{water} = 1.51$. The experimental threshold  for the atmospheric neutrino analysis is, however,  $E_\text{thresh} = 100$ MeV, which is higher than $\gamma_\text{water} m_e$ and it is what sets the threshold on the electron detectability. This energy threshold is set such as to avoid Michel electrons which are the electrons produced in muon decay \cite{michel}.
 \item \textit{Directionality}: As we have previously argued, signal electrons are emitted in the forward direction, and therefore are a good tool to point at the origin of BDM. The angular resolution of Super-K improves as a function of the electron energy up to a point where all photomultipliers saturate, in which case it degrades and it gets harder to infer the direction of the electron. We therefore take a conservative value of the angular resolution as $5^\circ$ across all energies studied ($E_e \in [100 \text{ MeV} - 100 \text{ GeV}]$). A more detailed study is required by the Super-K collaboration to find the appropriate resolution for this analysis. 
 This conservative resolution is smaller than the full extent of the GC in the sky, so it will not impact the results. For the dSphs searches, we only trigger on electrons within $5^\circ$ from a particular source location. 
 \item \textit{Gadolinium}: Gadolinium has one of the highest neutron capture rates. Tests have been conducted for its use in Super-K. When added to Super-K, gadolinium captures emitted neutrons in the $\overline{\nu_e} + p \rightarrow e^+ + n$ process and emits a distinctive 8 MeV photon, and therefore triggers on the $\overline{\nu_e} $ background \cite{Beacom:2003nk,Magro:aa,Mori:2013wua,Magro:2015sca,Xu:2016cfv,Nakahata:2015shm}. A full Super-K study will be able to estimate the reduction in background events when gadolinium is used, but it will not be included in this analysis.

\end{itemize}
Hyper-K is the future Super-K upgrade but with 25 times the fiducial mass,\footnotemark ~and thus will improve the sensitivity of Cherenkov detectors to BDM. In the following we assume it has the same properties as Super-K, from angular resolution to energy threshold \cite{Abe:2011ts,Kearns:2013lea,Abe:2014oxa,DiLodovico:2015kta,Lodovico:2015yii,Hadley:2016jpp}.

\subsection{Argon-Based Detectors: DUNE}

We now turn to the event selection at DUNE. DUNE is a planned LArTPC experiment which will be located at the Sanford Underground Research Lab. It will serve as the far detector for the long baseline neutrino facility and will be performing off-beam physics. It will include four 10-kton detectors. In the following, we study the sensitivity of 10 and 40 kton volume experiment to BDM \cite{Acciarri:2015uup}.
The BDM features that we use to select potential signal events are the following:
\begin{itemize}
\item \textit{Energy range}: To avoid being overwhelmed by the solar neutrino background, and to be conservative with the capability of the photodetector system to trigger on these events, we focus on the emitted electrons of energies $E_e > 30~\MeV$. This is a factor of 3 lower than a similar analysis at Super-K. Unlike Cherenkov detectors, Michel electrons are clearly associated with the parent muon track in LArTPCs. It is therefore easy to distinguish Michel electrons from electrons produced in charged current scatterings, and thus, the energy threshold can be lowered from 100 to 30 MeV.

\item \textit{Absence of hadronic processes}: The signal does not include any hadrons in the final state, and therefore, we can veto events with extra hadrons. The advantage of argon-based detectors over water/ice Cherenkov detectors is their ability to identify hadronic activity to low energies. We explore the details of the DUNE experiment in background discrimination in Sec. \ref{sec:background}.
\item \textit{Directionality}: A feature of the LArTPC technology is its good angular resolution. With an estimated $1^\circ$ resolution of low energy electrons, the DUNE experiment will be able to reduce the background for the dSphs searches as the search cone can be as small as the resolution. This resolution has been studied for energies $\mathcal{O}(1 ~\text{GeV})$, and further study from liquid argon experiments should be carried out for a more accurate value for sub-GeV electron energies.

\end{itemize}

\subsection{Detector summary}
\label{sec:detector}

\begin{table*}[t]
\begin{tabular}{l c c c c c}
\hline
\hline
Name & Number target $e^-$ & Energy Threshold & Angular Resolution & Exposure Time & Refs. \\
          &                                   & (MeV)   & (deg) & (years) \\
\hline
Super-K & $7.45 \times 10^{33}$  & 100 & 5 & 13.6 &  \cite{Fukuda:2002uc} \\
Hyper-K & $1.86 \times 10^{35}$ & 100 & 5 & 13.6 &  \cite{Abe:2011ts,Kearns:2013lea} \\
DUNE-10 kton  & $2.70 \times 10^{33} $ & 30 & 1 & 13.6 & \cite{Acciarri:2015uup} \\
DUNE-40 kton  & $1.08 \times 10^{34} $ & 30 & 1 & 13.6 & \cite{Acciarri:2015uup} \\
\hline
\end{tabular}
\caption{Detectors included in this analysis. We use the exposure time of Super-K as a reference for comparison with the rest of the experiments.} \label{tab:experiments}
\end{table*}

In \Tab{tab:experiments}, we summarize the experiments studied: Super-K and its upgrade Hyper-K for Cherenkov detectors, and two proposed volumes for DUNE as a LArTPC detector. Another detector with a potential of setting some limits on BDM is ICARUS  \cite{Arneodo:2001tx,Bueno:2003ei} as it ran 5 years deep underground with no cosmic contamination, but we expect Super-K with its present data set to set stronger limits on BDM. As a point of reference, we use the current Super-K exposure of 13.6 years for all experiments in order to estimate limits on BDM.

\footnotetext{The Hyper-K detector design might be modified for greater photomultiplier coverage and smaller mass \cite{hyperk}, but we assume the volume used in the initial letter of intent for this study \cite{Abe:2011ts}.}

\section{Background Modeling}
\label{sec:background}

We estimate the number of atmospheric neutrino background events in each experiment in turn. 
\subsection{Cherenkov Detectors}

For Super-K and by extension Hyper-K, atmospheric neutrino data are already available, and help estimate the number of neutrino background events expected per year. Since we are not provided the electron spectrum, we use the full data set of events shown in \Ref{Dziomba} as the background. We use the fully contained single-ring electron events over the four periods of Super-K, SK-I (1489 days), SK-II (798 days), SK-III (518 days) and SK-IV (1096 days), or for a total of 10.7 years. We estimate the number of background events per year over all energies (provided in two categories sub-GeV and multi-GeV events) to be
\be
\frac{N_\text{bkg}^\text{sky}}{\Delta T} = 923 ~\text{year}^{-1} \left( \frac{V_\text{exp}}{22.5~\text{kton}} \right),
\ee
where $V_\text{exp}$ is the experimental volume. 
The number of background events can be scaled up for estimates of Hyper-K. For the BDM search within a cone of angle $\theta_C$ around a source, the number of expected background events is then
\be
\frac{N_{\text{bkg}}^{\theta_C}}{\Delta T} = \frac{1- \cos \theta_C}{2}   \frac{N_\text{bkg}^\text{sky}}{\Delta T},
\ee
which in the case of the GC analysis\footnote{Although the optimal value for the opening angle of the search cone depends largely on the DM distribution (J-factor), it also depends on the angular distribution of the scattering process, and has to be optimized separately given a particular scattering.} and $\theta_C = 10^\circ$ is
\begin{eqnarray}
\frac{N_\text{bkg}^{10^\circ}}{\Delta T} &=&  7.0 ~\text{year}^{-1} \left( \frac{V_\text{exp}}{22.5~\text{kton}} \right).
\end{eqnarray}

A proper Super-K analysis can lower these estimates for the background by the use of the full background energy spectrum, and can thus improve the limits on BDM.

\begin{table*}[t]
\begin{tabular}{| c | c | c | c |}
\hline
\textbf{Final State Hadron} & \textbf{0 Produced ($\%$)}  & \textbf{1 Produced ($\%$)}  & \textbf{ $>$ 1 Produced ($\%$) } \\
\hline
p               & 17.7 & 50.4 & 31.8 \\
n               & 36.6 & 33.8 & 29.6 \\
$\pi^{\pm, 0}$  & 73.0 & 21.2 &  5.8 \\
$K^{\pm,0}$     & 99.4 & 0.5 &  0.1 \\
Heavier Hadrons & 98.9 &  1.1 &  0.00 \\
\hline
\end{tabular}
\caption{A summary of the production frequency of free hadrons in collisions between atmospheric electron (anti)neutrinos and argon-40.} \label{tab:hadronproduction}
\end{table*}

\subsection{LArTPC Detectors}

Previous studies have estimated the expected number of fully contained electron events to be 14053 per 350 kton year \cite{Acciarri:2015uup}. Therefore, we take the total number of electron events at DUNE to be 400 events per 10-kton-year.
 In order to optimize the analysis cuts, we generate a sample of 40,000 simulated atmospheric electron (anti)neutrino scattering events.
The reactions inside the pure $^{40}$Ar target volume are simulated using the GENIE neutrino Monte Carlo software (v2.10.6) \cite{Andreopoulos:2015wxa}. We model the atmospheric neutrinos with the Bartol atmospheric flux \cite{Gaisser:2002jj}. Since the flux varies slightly with geographic location and with altitude, we use the atmospheric flux available for the nearby MINOS far detector located in the Soudan Mine \cite{Barr:2004br}. We use the neutrino flux that occurs at solar maximum\footnote{One expects the most conservative limit to occur at solar minimum. Indeed the flux is higher at solar minimum, but it is dominated by lower energy neutrinos which produce pions. The detection threshold for pions is low enough to improve background rejection at this limit.} to provide the most conservative limit.
Although charged current processes dominate the background in this energy range, neutral current processes are also simulated. 

The dominant primary scattering processes are $\nu_e + n \rightarrow p + e^-$ and $\overline{\nu}_e + p \rightarrow n + e^+ $. However, due to secondary intranuclear processes the final observable state can, and generally will, include additional hadrons. These are comprised almost entirely of protons, neutrons, pions, and kaons. Table \ref{tab:hadronproduction} summarizes the frequency of different hadrons to be produced in the final state.

Approximately 99.72$\%$ of the simulated interactions contain a free hadron in the final state. This is a useful discriminant as a DM event would not produce a hadron in the final state. So, contingent on detectability, we are able to use these hadrons as a veto on charged current events.

\begin{table}[h]
\centering
\begin{tabular}{| c | c |}
\hline
\textbf{Hadron} & \textbf{Detection Threshold (MeV)} \\
\hline
p & 21 \\
$\pi^{\pm, 0}$  & 10 \\
$K^{\pm,0}$     & 17 \\
\hline
\end{tabular}
\caption{Kinetic energy thresholds for DUNE to be able to detect various hadrons \cite{Xin}.} \label{tab:hadronthresh}
\end{table}

To detect the emitted hadrons, DUNE is able to resolve hadronic activity down to low energy thresholds, provided in Table \ref{tab:hadronthresh} \cite{Xin}. 
Neutrons are harder to detect, and to be conservative, we assume that all neutrons escape detection, although future simulations of argon detectors might prove otherwise.
Implementing the hadronic veto to the simulated dataset, we find that less than $32\%$ of simulated background processes pass the cut based on hadron tagging alone. We therefore estimate the number of background events over the whole sky to be 
\be
N_{\text{bkg}}^{\text{all sky}} = 128 ~\text{events}/\text{year} \left( \frac{V_\text{exp}}{10~ \text{kton}}\right).
\ee
For the searches within $10^\circ$ around the GC, the number of background events is
\begin{eqnarray}
N_{\text{bkg}}^{10^\circ} &=& 1.0 ~\text{events}/\text{year} \left( \frac{V_\text{exp}}{10 ~\text{kton}}\right).
\end{eqnarray}
Using the angular information of the events found by looking up to 10 degrees around the DM sources such as the GC, the background is about 1 event per year.

\section{Reach at neutrino experiments}
\label{sec:dune}

\begin{figure}[t!]
\begin{center}
\includegraphics[width=18pc, trim = 0 0 0 0]{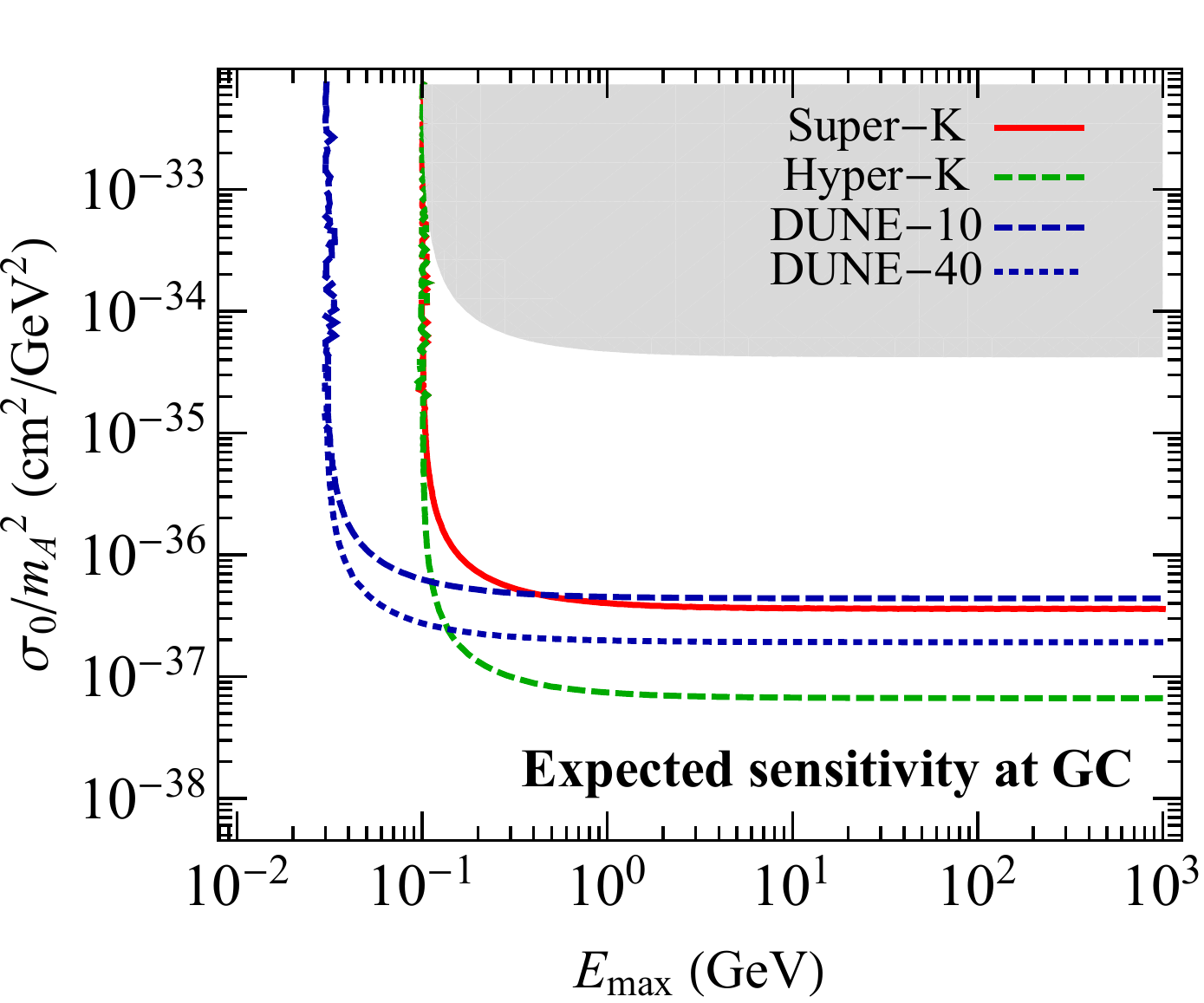}\hspace{2pc}%
\caption{\label{fig:sig_low}$95\%$ limits on parameter space for BDM annihilation for Super-K, Hyper-K and DUNE for 10 and 40 kton in volume. The gray region is excluded by the fact that no excess has been detected in Super-K in the past 11 year data set. }
\end{center}
\end{figure}

We now estimate the experimental sensitivity for BDM searches in the GC, leaving the analysis of dSphs to \Sec{sec:dSphs_analysis}. We compare Cherenkov detectors' large volume with the LArTPC's ability to reduce background events through particle identification and explore key experimental features such as low energy thresholds and excellent angular resolution for both technologies.

To measure the sensitivity of an experiment, we define the signal significance as 
\be \label{eq:sensitivity}
\text{Significance} = \frac{S}{ \sqrt{S +B }},
\ee
where $S$ is the number of signal events, and $B$ the number of background events. In the following, we estimate limits on the region of parameter space defined in \Sec{sec:detector} for a $2 \sigma$ significance, using the exposure time shown in \Tab{tab:experiments}.

In \Fig{fig:sig_low}, we show the $95\%$ limits of Super-K, Hyper-K, and DUNE to the effective cross section $\sigma_0$, defined in \Eq{eq:sigma0}, as a function of $E_\text{max}$, defined in \Eq{eq:emax}, in the constant amplitude limit.
In the case of light BDM ($2 m_e E_B \gg m_B^2$), $E_\text{max} \approx E_B$, while in the case of a heavy BDM ($2 m_e E_B \ll m_B^2$), $E_\text{max} \approx 2 m_e \gamma_B^2 $. We plot the combination  $\sigma_0/m_A^2$ since the number of signal events scales with the number density squared of DM in the case of annihilation. 

We also show in \Fig{fig:sig_low} as the gray region, the bounds set currently by Super-K without any angular information, having assumed a systematic deviation in the number of events $\delta N_\text{bkgd}/ N_\text{bkgd} = 10 \%$. 
This excludes cross sections per mass squared above $\sim 10^{-34} \text{cm}^2/\text{GeV}^2$.
We find that DUNE with 10 kton is almost equally sensitive to BDM signals as Super-K is, for the same exposure, even though DUNE is three times smaller. This is due to its improved background rejection. DUNE can also explore lower electron energies at a comparable angular resolution and therefore lighter BDM. 

Although different detector technologies can probe different features, DUNE can test for lighter BDM, while Super-K/Hyper-K can explore lower cross sections due to their large volumes.  It is crucial that there is an overlapping region between both experiments; it allows the two experiments to cross-check possible signals and limits, which is especially interesting when comparing different technologies. Detecting a signal in both experiments would be one step towards confirming a DM detection. 

\section{Dwarf Spheroidal Analysis}\label{sec:dSphs_analysis}

\begin{figure}[t]
\begin{center}
\includegraphics[width=18pc, trim = 0 0 0 0]{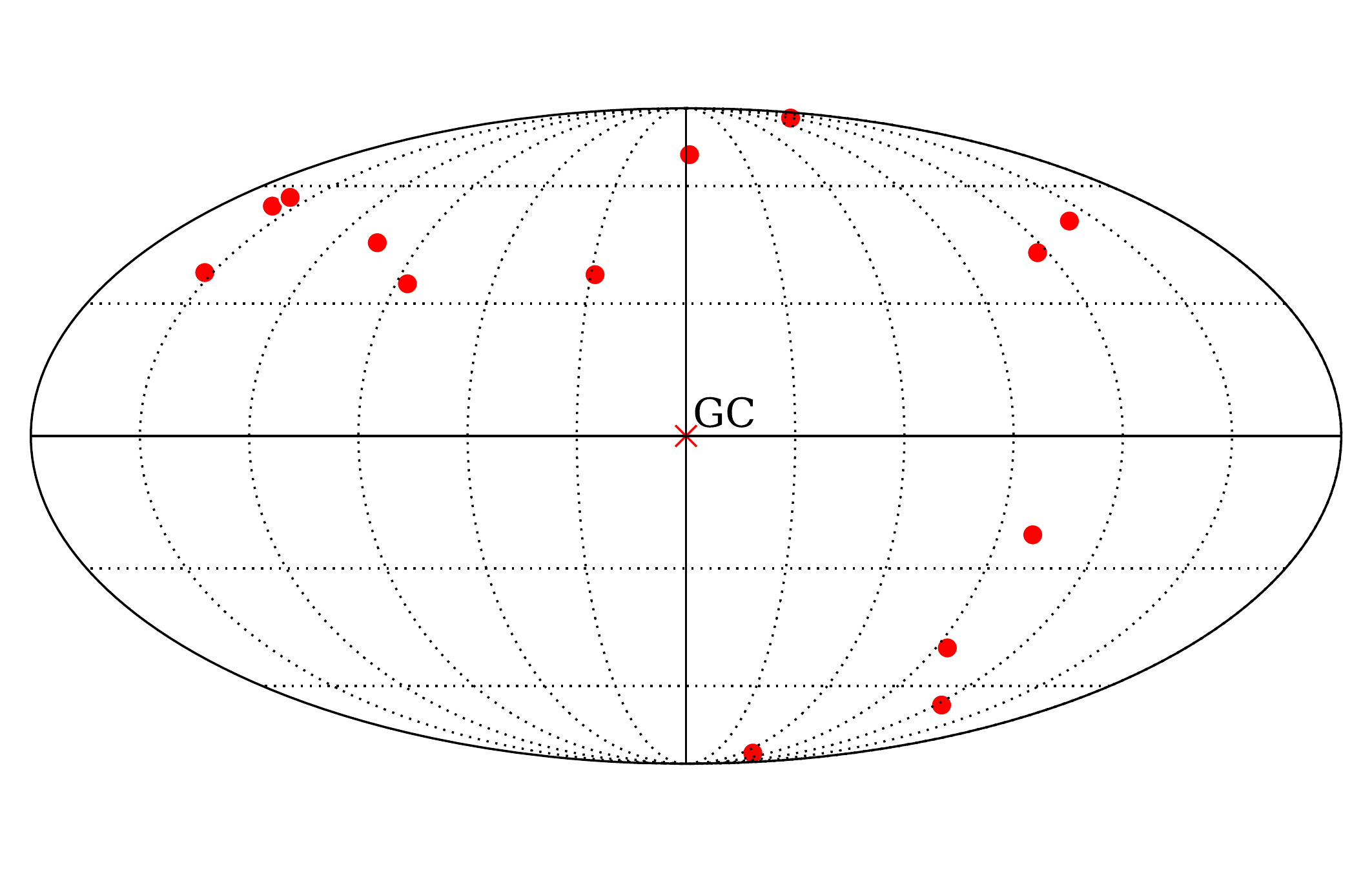}\hspace{2pc}%
\caption{\label{fig:dSphsMap} Map of the dSphs' locations in Galactic coordinates used in this analysis. The center of the figure is the GC.}
\end{center}
\end{figure}

Dwarf spheroidals are Milky Way satellite galaxies which are dense in DM and low in baryons; they are therefore good candidates for indirect detection searches, with low backgrounds \cite{Mateo:1998wg,McConnachie:2012vd}.
Although dSphs are less dense in DM than the GC, we can increase the sensitivity to BDM by stacking dSphs. In order to do so, we plot the direction of detected electron events in galactic coordinates, and correlate them with known sources within the experimental angular resolution, such as the dSphs as shown in \Fig{fig:dSphsMap}.

\subsection{J-factor of Dwarf Galaxies} \label{sec:dSphs}

Over the past few years, many dSphs have been found in large surveys \cite{Koposov:2015cua,Bechtol:2015cbp}. We list in \Tab{tab:dsphs} the locations of the brightest dSphs (in J-factors), the separating distance from the Earth, as well as their found J-factors in decay and annihilation, assuming a NFW profile.

The J-factors listed are integrated over a cone of half angle $0.5^\circ$ due to their small extent in the sky. Therefore, in detecting these sources, the search cone (see \Fig{fig:cone}) has to be as small as possible and we therefore choose it to be the experimental angular resolution.

\begin{table*}[t]
\begin{tabular}{ l c c c c c c}
\hline
\hline
  Name & \textit{l} & $b$ & Distance (kpc) & $\log_{10} (J_\text{ann}) $& $\log_{10} (J_\text{dec}) $ & Refs.  \\
 &  (deg) & (deg)&(kpc)&($\log_\text{10}$ [GeV$^2$ cm$^{-5}$]) & ($\log_\text{10}$ [GeV cm$^{-2}$])  &  \\
\hline
Bootes I & 358.1 & 69.6 & 66 & $18.8 \pm0.22 $ & $17.9 \pm 0.26$  & \cite{DallOra:2006pt}\\
Carina & 260.1 & -22.2 & 105 & $18.1 \pm 0.23$ & $17.9 \pm 0.17$ & \cite{Walker:2008ax}\\
Coma Berenices & 241.9 & 83.6 & 44 & $19.0 \pm 0.25 $ & $18.0 \pm 0.25$ & \cite{Simon:2007dq}\\  
Draco & 86.4 & 34.7 & 76 & $18.8 \pm 0.16 $ & $18.5 \pm 0.12 $ &\cite{Munoz:2005be}\\
Fornax & 237.1 & -65.7 & 147 & $18.2 \pm 0.21 $ & $17.9 \pm 0.05$ & \cite{Walker:2008ax} \\
Hercules & 28.7 & 36.9 & 132 & $18.1 \pm 0.25$ & $16.7 \pm 0.42$ & \cite{Simon:2007dq} \\
Reticulum II & 265.9 & -49.6 & 32 & $19.6 \pm 1.0$ & $18.8 \pm 0.7$ & \cite{Bonnivard:2015tta,Bechtol:2015cbp} \\
Sculptor & 287.5 & -83.2 & 86 & $18.6 \pm 0.18$ & $18.2 \pm 0.07$ & \cite{Walker:2008ax} \\ 
Segue 1  & 220.5 & 50.4 & 23& $19.5 \pm0.29$ & $18.0 \pm0.31$ & \cite{Simon:2010ek} \\
Sextans & 243.5 & 42.3 & 86 & $18.4 \pm 0.27$ & $17.9 \pm 0.23$ & \cite{Walker:2008ax} \\
Ursa Major I & 159.4 & 54.4 & 97 & $18.3 \pm 0.24$ & $17.6 \pm0.38$ & \cite{Simon:2007dq,Geringer-Sameth:2014yza} \\
Ursa Major II & 152.5 & 37.4 & 32 & $19.3 \pm 0.28$ & $18.4 \pm 0.27$ &\cite{Simon:2007dq} \\
Ursa Minor & 105.0 & 44.8 & 76 & $18.8 \pm 0.19$ & $18.0 \pm 0.16$ & \cite{Munoz:2005be} \\
Willman 1 & 158.6 & 56.8 & 38 & $19.1 \pm 0.31$ & $17.5 \pm 0.84$ & \cite{Willman:2010gy,Essig:2009jx}\\
\hline
\end{tabular}
\caption{Table of dSphs's locations, distances and J-factors, compiled in Refs. \cite{Ackermann:2015zua,Geringer-Sameth:2014yza}}The decay J-factors were taken from Ref. \cite{Geringer-Sameth:2014yza} assuming the largest error. \label{tab:dsphs}
\end{table*}

Although individually the J-factors of dSphs are 2 orders of magnitude lower than that of the GC, one can perform a stacked analysis of the dSphs which would effectively sum over the J-factors of all the dSphs considered to set more constraining limits. Such analysis is interesting as it can be a confirmation that a signal is potentially that of DM if it is detected in both the GC and dSphs. 

\subsection{Event Reach}

We compute the number of background events as in \Sec{sec:background}, but here we limit the search angle to the experimental resolution. We find
\begin{eqnarray}
\frac{N_\text{bkg}^{5^\circ}}{\Delta T} &=& N_\text{dSphs} ~1.8 ~\text{year}^{-1} \left( \frac{N_\text{target}}{7.45 \times 10^{33}} \right), \nonumber \\
&& \qquad  \qquad ~~~ \text{for Super-K} \\
\frac{N_\text{bkg}^{1^\circ}}{\Delta T} &=&  N_\text{dSphs} ~0.01 ~\text{year}^{-1} \left( \frac{N_\text{target}}{2.70 \times 10^{33}} \right), \nonumber \\
&& \qquad \qquad ~~~ \text{for DUNE}
\end{eqnarray}
where $N_\text{dSphs}$ is the number of dSphs considered in the analysis.

\begin{figure}[t]
\begin{center}
\includegraphics[width=18pc, trim = 0 0 0 0]{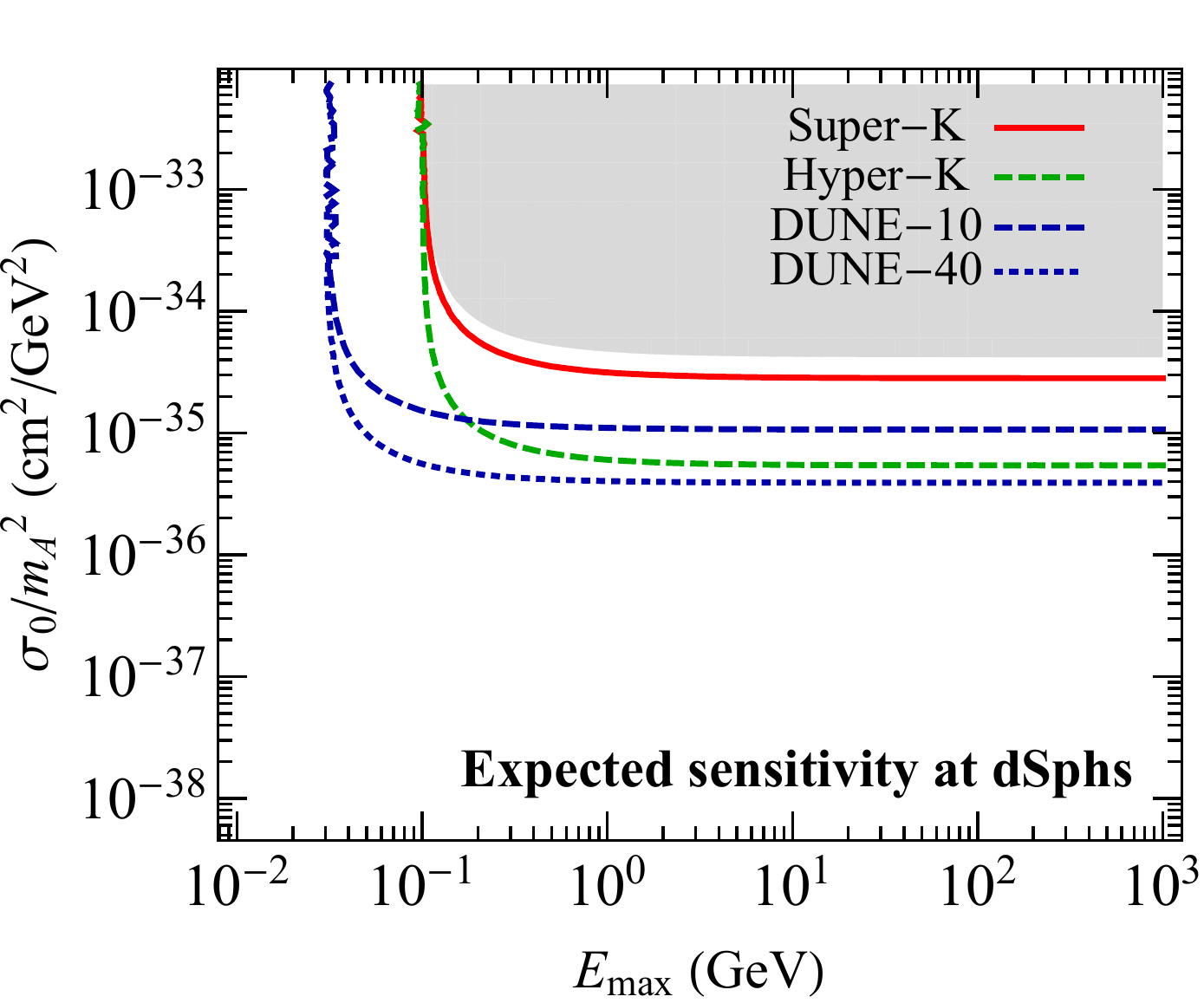}\hspace{2pc}%
\caption{\label{fig:sig_high_dwarf} $95\%$ limits on parameter space for BDM annihilation in a stacked analysis of dSphs. The gray region is excluded by the fact that no excess has been detected in Super-K in the past 11 year data set.}
\end{center}
\end{figure}

Similarly to the GC analysis, we show in \Fig{fig:sig_high_dwarf} the different experimental sensitivities. Although the reach is not as deep as that of the GC analysis, the dSphs analysis would be an excellent confirmation that any potential signal found in the GC is indeed consistent with a DM interpretation. Also, with future surveys, one might be able to push further the dSphs analysis sensitivity by finding more dSphs.

We also point out in this analysis that DUNE with only 10-kton will be able to outperform Super-K due to its excellent background rejection enabled by $1^\circ$ angular resolution. One caveat of this analysis is that when reducing the search cone to only 1 degree and 5 degrees for DUNE and Super-K respectively, we are only able to set limits reliably on BDM with a high boost factor $\gamma_B$ as the events have to be extremely forward (see \App{app:forward}).

\section{Conclusions}
\label{sec:conclusion}

In this work, we have studied the experimental signatures of a class of DM models called boosted dark matter, in which one component has acquired a large Lorentz boost today and can scatter off electrons in neutrino experiments. Our analysis compared two neutrino technologies: Liquid argon detectors like DUNE and Cherenkov detectors like Super-K and Hyper-K.

We compared the excellent particle identification of LArTPC detectors by simulating neutrino events in argon, with the large volume of Cherenkov experiments to help further reduce the atmospheric neutrino background. Building a search strategy tuned for each experiment extends the physics reach of neutrino detectors from classic DM indirect detection to BDM direct detection, enabled by the ability to tag BDM particle on almost an event-by-event basis, especially in liquid argon experiments. 

If the BDM component has a much higher energy than the electron mass, the electron is emitted in the forward direction, and can thus be used to trace back the origin of DM. Such a feature, coupled with a good angular resolution in neutrino experiments can help establish limits on BDM. The angular resolution can also help point back to the origin of DM; constructing a map of the origin of these sources can help correlate signals from neutrino detectors with other experiments, for example gamma rays at Fermi \cite{Atwood:2009ez}. 

If a signal is detected, some BDM properties can be extracted. For example, the maximum Lorentz boost for an electron is related to that of the $B$ particle by
\be 
\gamma_e^{\text{max}} = 2 \gamma_B^2 - 1. \label{eq:gamma}
\ee
We can therefore extract $E_e^{\text{max}}$ from the electron spectrum and obtain the boost factor of $B$. In the case of a monoenergetic signal, where all particles $B$ have energy $E_B$, we obtain a single value of $\gamma_B$. As we expect low statistics, we can only bound the Lorentz factor from below.

We performed two analyses, one for BDM originating from the GC and one in which we stacked signals from dSphs. We found that DUNE with 10 kton can perform as well as Super-K in the case of the GC analysis, and can outperform it in the dSphs analysis due to its superior angular resolution. 
In both analyses, we adopted a conservative strategy, in particular by using all atmospheric data across a wide range of energies as background. A dedicated experimental search from the Super-K and DUNE collaborations is able to properly estimate the background and improve the limits on BDM.

The largest constraints affecting the parameter space studied in this work are from our analysis of published Super-K data, where Super-K has not detected any excess of electron events over muon events above statistical fluctuations. 
Such limits are set without any angular information, and thus can be extended by the Super-K, Hyper-K and DUNE collaborations through a similar analysis to the one described in this work.
 Other limits, although not discussed above, are model specific and need to be taken into consideration when building a BDM model. 
 These limits include direct detection bounds on any thermal component of a particle interacting with electrons and/or quarks: Direct detection limits on electron scattering are set by the same process that enables the $B$ particle detection at neutrino experiments, but affect the thermal $B$ component instead of the relativistic one \cite{Essig:2012yx}. Direct detection limits from proton scattering would affect $B$ particles with masses larger than $\mathcal{O}(1)$ GeV, making the ability of the DUNE experiment to lower the energy detection threshold of utmost importance \cite{Barreto:2011zu,Agnese:2013jaa,Akerib:2015rjg}. 
 Other possible limits include cosmic microwave background (CMB) constraints on the power injected by the thermal $B$ component into SM particles at early redshifts \cite{Madhavacheril:2013cna}. All these limits need to be studied properly when discussing a particular model of BDM. An example of such study has been implemented in \Ref{Agashe:2014yua}.

DUNE is an excellent detector to cross-check present Cherenkov detectors and extend the reach of neutrino detectors in DM searches. 
Having multiple technologies for the hunt of DM is key in its eventual detection.

\section*{Acknowledgements}

We thank Hongwan Liu, Nicholas Rodd and Jesse Thaler for notes on the manuscript. We also thank Kaustubh Agashe, Jonathan Asaadi, Daniel Cherdack, Gabriel Collin, Yanou Cui, Hugh Gallagher, Chris Kachulis, Ed Kearns, Ian Moult, and Jesse Thaler for helpful discussions.
L.N. is supported by the U.S. Department of Energy under cooperative research agreement contract number DE$-$SC00012567. JMC and JM are funded through NSF grant 1505855. TW is supported by the Pappalardo Fellowship Program at MIT.

\appendix

\section{Understanding Forward Scattering}\label{app:forward}

In \Sec{sec:detection}, we assumed that when the energy $E_B$ of the boosted particle is greater than the electron mass  
\be \label{eq:ebggme}
E_B \gg m_e,
\ee
 the final state electron of the elastic scattering $B e^- \rightarrow B e^-$ is emitted in the forward direction. This is crucial as the observed electron can then point back to the origin of the $B$ particle. From kinematics, the scattering angle of the emitted electron relative to the incoming $B$, labeled $\theta_e'$ as shown in \Fig{fig:cone}, is
\be \label{eq:cos}
\cos \theta'_e = \frac{E_B + m_e}{\sqrt{E_B^2 - m_B^2}} \sqrt{\frac{E_e - m_e}{E_e + m_e}},
\ee
where the energy of the emitted electron is $E_e$. Applying the assumption of \Eq{eq:ebggme}, \Eq{eq:cos} becomes
\begin{eqnarray} \label{eq:costheta}
\cos \theta'_e &=& \sqrt{\frac{1-1/\gamma_e}{1+1/\gamma_e}} \frac{\gamma_B}{\sqrt{\gamma_B^2 - 1}} \nonumber \\
& \approx&  \left(1 - \frac{1}{\gamma_e} \right) \left(1 + \frac{1 }{2 \gamma_B^2} \right) + \mathcal{O} \left(\frac{1}{\gamma_e^2}, \frac{1}{\gamma_B^4} \right), \nonumber \\
\end{eqnarray}
where 
\be
\gamma_i = E_i/m_i
\ee
with $i \in \{ B, e \}$ being the $B$ and electron boost factors.
We have expanded in large $\gamma_e$ and $\gamma_B$ in \Eq{eq:costheta}.

In the cases where $\gamma_B, \gamma_e \gg 1$, we find to a good approximation that $\cos \theta'_e \approx 1$ and $\sin \theta'_e \approx 0$. The angle of the recoiled electron relative to the DM source $\theta_e$ is related to $\theta'_e$ by
\be \label{eq:costhetae}
\cos \theta_e = \cos \theta_B \cos \theta'_e - \sin \theta_B \sin \phi'_e \sin \theta'_e \stackrel{\theta_B \rightarrow 0}{\approx} \cos \theta'_e,
\ee
where $\phi'_e$ is the azimuthal angle of the recoiled electron with respect to the incoming $B$ as shown in \Fig{fig:cone} and is uniformly distributed between $0$ and $2 \pi$.

In order to estimate the error on the measured angle $\theta_e$ compared to the incoming $B$ angle $\theta_B$, we study the deviations in \Eq{eq:costhetae} from $\cos \theta_e = \cos \theta_B$. Taylor expanding around $\theta'_e  = 0$, we find 

\begin{eqnarray}
\cos \theta_e &=& \cos \theta_B  - \theta'_e \sin \theta_B \sin \phi'_e   + \mathcal{O} ((\theta'_e)^2). 
\end{eqnarray}
From \Eq{eq:costheta}, and in terms of the boost factors $\gamma_e$ and $\gamma_B$,
\be
\theta'_e \approx \sqrt{2} \left( \frac{1}{\gamma_e} -\frac{1}{2 \gamma_B^2}  \right)^{1/2} + \mathcal{O} (1/\gamma_e^2, 1/\gamma_B^4) \approx \sqrt{2/\gamma_e}.
\ee
The last approximation is found from the kinematics relation $\gamma_e^\text{max} = 2 \gamma_B^2 - 1 $ and therefore $\gamma_e < 2 \gamma_B^2$.
Taking $\sin \phi'_e = 1$ as its maximum value, we find that the deviation from the forward approximation is
\begin{eqnarray}
\cos \theta_e &=& \cos \theta_B  - \sqrt{2/\gamma_e} \sin \theta_B  
\end{eqnarray}
\begin{figure}[t]
\begin{center}
\includegraphics[width=15pc, trim = 0 0 0 0]{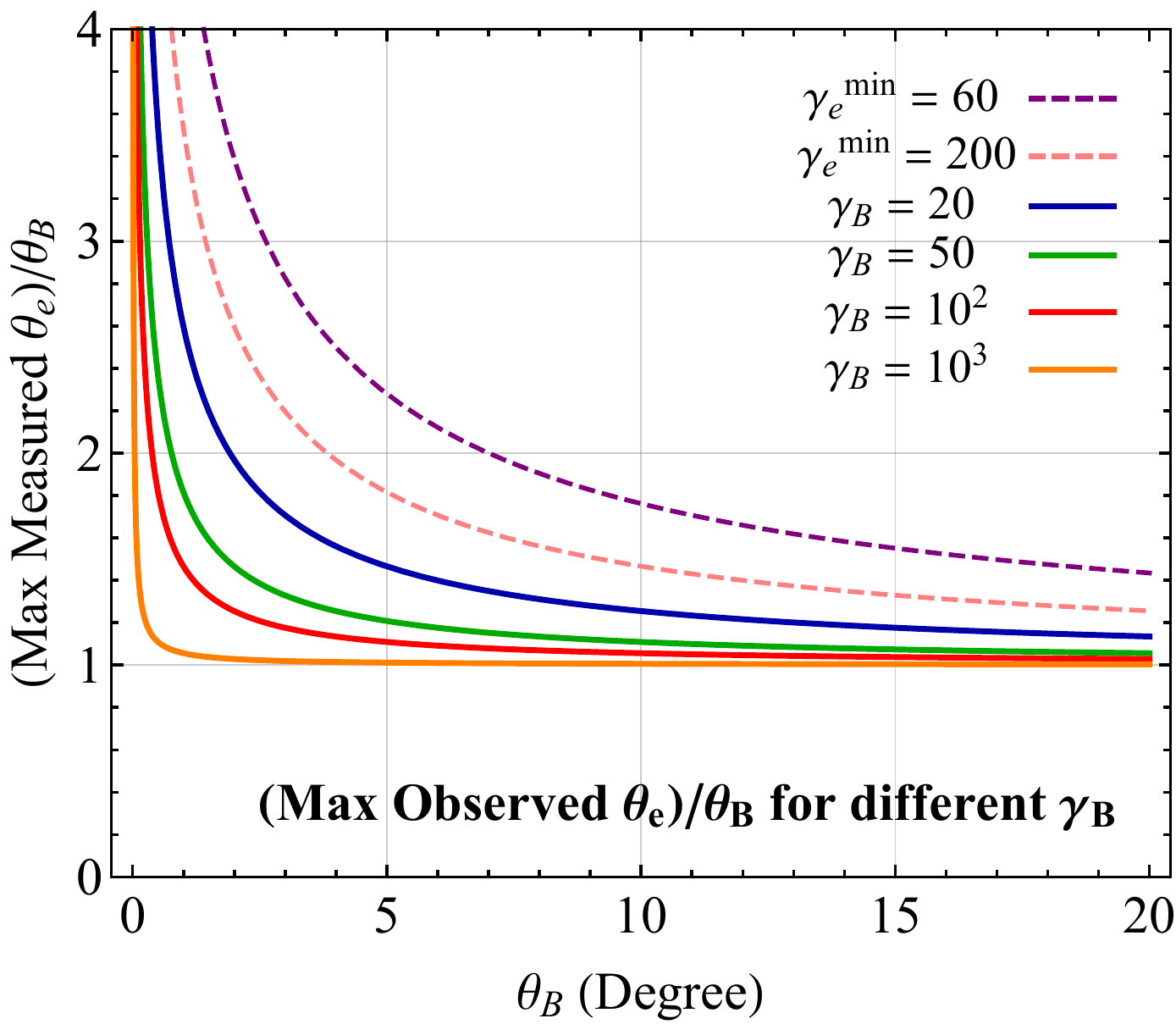}\hspace{2pc}%
\caption{\label{fig:angularerror} Maximum observed angle of the electron $\theta_e$ as a function of the initial angle at which the Boosted particle $B$ was emitted for different values of the boost factor $\gamma_B$.  }
\end{center}
\end{figure}

We show the results of the ratio of the observed electron angle by the incoming $B$ angle $\theta_e/\theta_B$, as a function of the $B$ angle $\theta_B$ in \Fig{fig:angularerror}. 
For every value of $\gamma_e$ found, there exists a minimal gamma factor of the original particle $B$ such that $\gamma_e = 2 (\gamma_B^\text{min})^2 -1 $. The solid curves in \Fig{fig:angularerror} correspond to the ratio $\theta_e/\theta_B$ with
\begin{eqnarray}
\theta_e &=& \arccos( \cos \theta_B - \sqrt{2/\gamma_e} \sin \theta_B) \nonumber \\
&=& \arccos( \cos \theta_B - \sqrt{2/(2(\gamma_B^\text{min})^2 -1 )} \sin \theta_B), \nonumber \\
\end{eqnarray}
for different values of $\gamma_B^\text{min}$. We find that values of $\gamma_B > 20$ are suitable within the forward scattering approximation, with errors less than $20 \%$. 
We also study the largest value of $1/\gamma_e$, which occurs at the experimental threshold $E_\text{thresh}$
\be
\gamma_e^\text{min} = E_\text{thresh}/m_e.
\ee
 As discussed in \Sec{sec:detection}, the experiment thresholds considered are $E_\text{thresh} = 30$ MeV and $E_\text{thresh} = 100$ MeV, which lead to a gamma factor of $\gamma_e^\text{min} = 60 - 200$. We show the measured angle of the electron off the source as a function of the initial BDM angle $\theta_B$ for the events right at the energy threshold in dashed lines in \Fig{fig:angularerror}. This study can be properly incorporated within the experimental framework to estimate the systematics as a function of the emitted electron's energy.

\section{Comparing the Full Analysis with a Concrete Model}
\label{sec:bdm}

\begin{figure*}[t]
\begin{center}
\includegraphics[scale=0.6, trim = 0 0 0 0]{plots/AABB}\hspace{2pc}%
\includegraphics[scale=0.6, trim = 0 0 0 0]{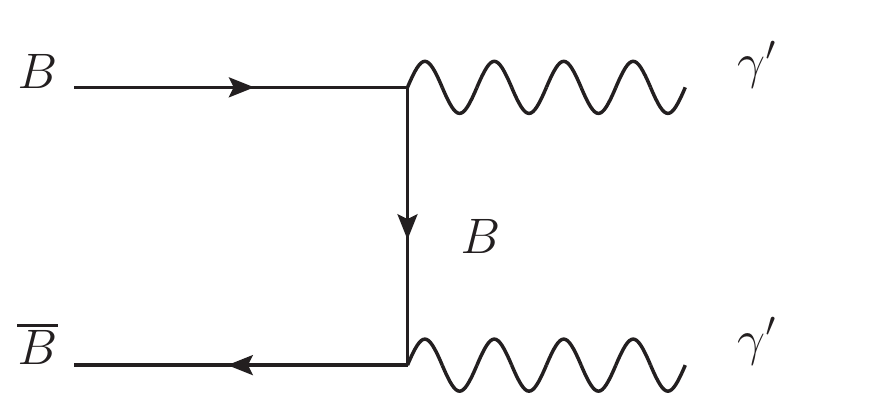}\hspace{2pc}%
\includegraphics[scale=0.5, trim = 0 0 0 0]{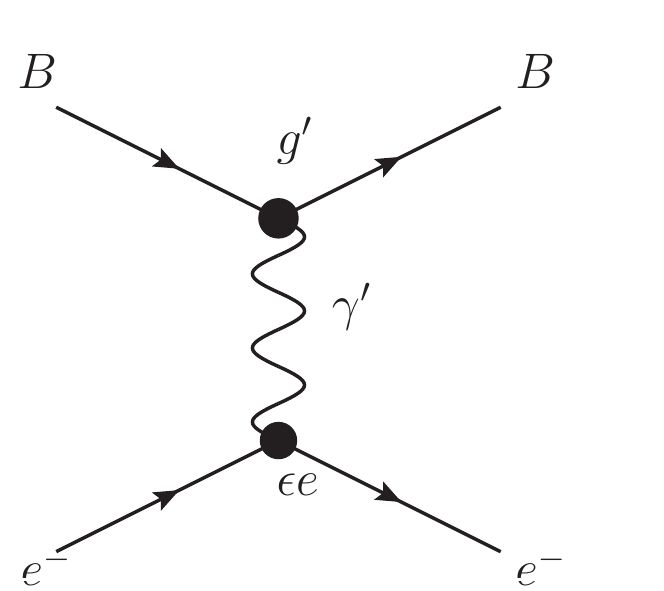}\hspace{2pc}%
\caption{\label{fig:diagrams} Feynman diagrams for the production and detection of DM particles. (Left) Diagram that controls the abundance of $A$ in the early universe as well as today's production of $B$ with a Lorentz boost through $A$ annihilation. (Middle) Annihilation of $B$ to $\gamma'$, diagram that contributes to CMB limits. (Right) Signal diagram of $B$ scattering off electrons.}
\end{center}
\end{figure*}

In this section, we summarize the model explored in \Ref{Agashe:2014yua}, based on \Ref{Belanger:2011ww}, and show the reach of the DUNE experiments in the appropriate parameter space.
We start with a multicomponent DM model with two particle species $A$ and $B$, such that $A$ is the dominant DM component that interacts solely with $B$, and $B$ is the subdominant component that couples to the standard model. If a mass hierarchy exists such that $m_A \gg m_B$, the annihilation process $A \overline{A} \rightarrow B \overline{B}$ leads to particles $B$'s with energies $E_B = m_A$ and thus a high boost factor $\gamma_B = m_A/m_B$. 

We further take the $B$-SM couplings to be through the kinetic mixing of a dark photon $\gamma'$ with the photon. The mixing term is
\be
\mathcal{L} \supset -\frac{\epsilon}{2} F'_{\mu \nu} F^{\mu \nu},
\ee
where $F'^{\mu \nu}$ is the dark photon field, $F^{\mu \nu}$ is the photon field, and $\epsilon$ is the coupling of the interaction. We take the coupling of $B$ to the dark photon to be $g'$, which is large but perturbative. The model parameters are therefore:
\be
m_A , ~~ m_B, ~~ m_{\gamma'}, ~~ g', ~~ \epsilon.
\ee

The cross section of the $A - \overline{A}$ annihilation (see the left diagram of \Fig{fig:diagrams}) is set such that we obtain the right abundance of $A$'s today, which brings the value of the cross section close to the thermal cross section. The abundance of $B$ particles is controlled by both the annihilation of the $A$ diagram as well as the annihilation of  the $B$ diagram (middle diagram of Fig. \ref{fig:diagrams}).

Finally, the scattering of $B$ particles off electrons is set by the right diagram of Fig. \ref{fig:diagrams}. The same diagram with a nucleon instead of an electron is the one that sets direct detection bounds on the thermal component of $B$. This study focuses however on $B$ particles with masses below the ones studied so far in direct detection experiments. Of course higher $B$ masses can be evaded by the introduction of inelastic scattering \cite{TuckerSmith:2001hy,Cui:2009xq}.

\begin{figure}[t]
\begin{center}
\includegraphics[width=20pc, trim = 0 0 0 0]{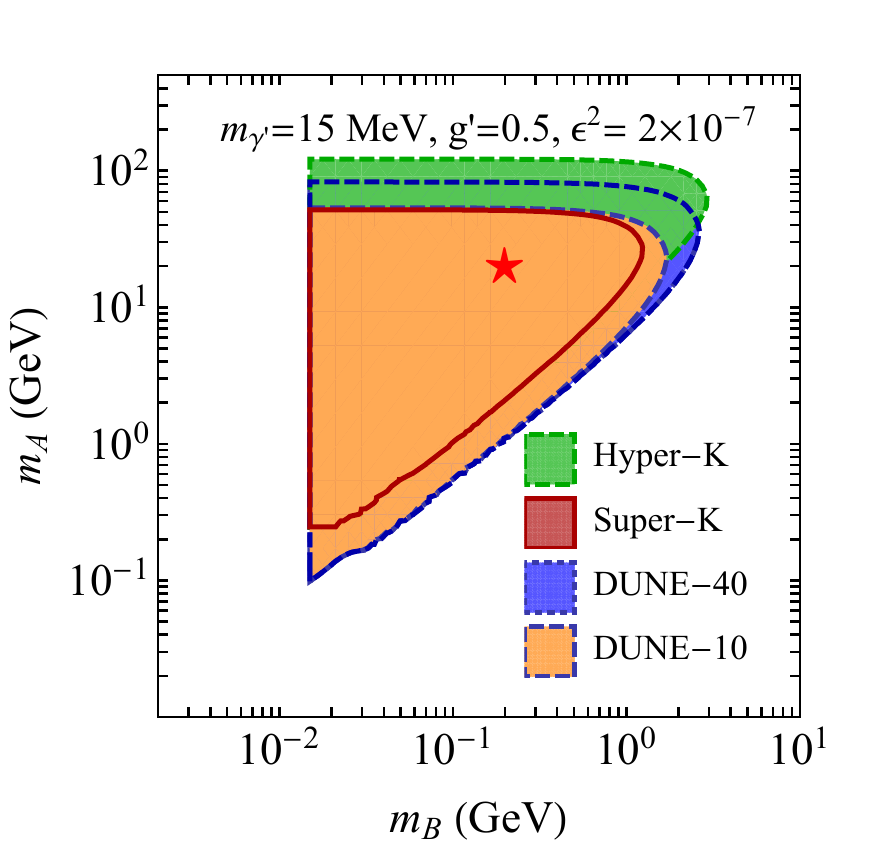}\hspace{2pc}%
\caption{\label{fig:old_model} Super-K, Hyper-K and DUNE limits for the model from \Ref{Agashe:2014yua} with an exposure of 13.6 years. }
\end{center}
\end{figure}

For Fig. \ref{fig:old_model}, we use the following benchmark (while varying $m_A$ and $m_B$), where the limits on the dark photon are consistent with those in Ref.  \cite{Goudzovski:2014rwa}. 
\be \label{eq:benchmark}
 \qquad m_{\gamma'} = 15 ~\text{MeV},~~ g' = 0.5,~~  \epsilon^2  = 2 \times 10^{-7}. 
\ee

In Fig. \ref{fig:old_model}, we show the estimated limits of DUNE as well as Super-K and Hyper-K in the $m_A - m_B$ space, first presented in \Ref{Agashe:2014yua}. We find consistent results with Fig. \ref{fig:sig_low}, as Super-K and DUNE with 10 kton have similar sensitivity, with DUNE able to probe lower electron recoils. This is shown by the diagonal line in the triangular range of Fig. \ref{fig:old_model} which can be thought of as the difference between $m_A$ and $m_B$, a quantity that is related to the energy of the emitted electron. 

\bibliography{biblio}
\end{document}